\documentclass[twocolumn]{jpsj3}
\usepackage{txfonts}
\title{Dimensional Dependence of Critical Exponent of the Anderson Transition in the Orthogonal Universality Class}

\author{\name{Yoshiki \surname{Ueoka}}, \name{Keith \surname{Slevin}}}
\inst{Department of Physics, Graduate School of Science, Osaka University, 1-1 Machikaneyama, Toyonaka, Osaka 560-0043, Japan}

\abst{We report improved numerical estimates of the critical exponent of the Anderson transition in Anderson's model of localization in $d=4$ and $d=5$ dimensions.
We also report a new Borel-Pad\'e analysis of existing $\epsilon$ expansion results that incorporates the asymptotic behaviour for $d\to \infty$ and gives better agreement with available numerical results.}

\kword{Anderson transition, transfer matrix, finite size scaling, kicked rotor model, nonlinear $\sigma$ model, \\ $\epsilon$ expansion, Borel-Pad\'e analysis}

\begin{document}
\maketitle
\newpage
\section{Introduction}

The classical motion of an electron in a random potential is diffusive. Anderson\cite{anderson58} realized that diffusion may be
suppressed in the corresponding quantum problem.
This phenomenon was subsequently called Anderson localisation.
The suppression of diffusion is associated with a metal-insulator transition called the Anderson transition.
The Anderson transition is a continuous quantum phase transition and the renormalization group, developed
to describe continuous thermal phase transitions, is applicable\cite{abrahams79}.

One feature of continuous phase transitions is the power law dependence of various physical quantities, called critical phenomena, observed near the
transition\cite{nishimori11}.
The exponents appearing in these laws are expected to be universal depending only on the symmetries of the Hamiltonian and the dimensionality $d$ of the system.
Two important symmetries for the Anderson transition are time reversal symmetry and spin rotation symmetry (see Ref. \citen{evers08} for a full discussion of relevant symmetries and universality classes).

In this paper we are concerned with systems which have both time reversal symmetry and spin rotation symmetries.
Such systems are said to have orthogonal symmetry.
For systems with this symmetry the lower critical dimension is $d=2$.
In $d=1$ and $d=2$, arbitrarily weak randomness is enough to cause localisation.
While, in $d>2$, localization only occurs when the randomness or disorder is sufficiently strong.
Our focus is on the dimensionality dependence of the critical exponent $\nu$ that describes the power law divergence of the correlation length around the critical point.

One motivation for our work is that experimental realisations of  higher dimensional Anderson transitions are now
foreseeable.
Such realisations will probably not involve electrons, where, in any case, there has been only very slow progress in understanding how
the Coulomb interaction between electrons affects the transition.\cite{harashima12,Burmistrov13,amini14,harashima14}.
Rather, they may be possible in systems that display dynamical localisation.

Dynamical localisation is the analogue of Anderson localisation in momentum space for periodically driven systems.
One such a system is the one dimensional quantum kicked rotor, which was experimentally realised by Moore et al\cite{moore95}.
In Ref. \citen{casati89}, a mapping between a one dimensional kicked rotor where the amplitude of the periodic kick is modulated in an aperiodic manner controlled by three incommensurate frequencies and a three dimensional lattice model was derived.
A realisation of this system in a cold atomic gas was reported in Ref. \citen{lemarie08} where
a transition driven by the kicking strength between a phase of diffusion and localisation
in momentum space was observed.
A scaling analysis of the experimental data yielded a value $\nu=1.4 \pm .3$ for the critical exponent.
Numerical simulations of the experimental system, reported in the same reference, gave $\nu=1.60 \pm .05$.
In subsequent work the precision of the experimental measurement of the critical exponent has
improved considerably with a value $\nu=1.63 \pm .05$ reported in Ref. \citen{lopez12}.
These values are in  agreement with the most precise numerical estimates $\nu = 1.571 \pm .004$\cite{slevin99a,slevin14} and $\nu = 1.590 \pm .006$\cite{rodriguez10,rodriguez11} of the critical exponent
for the $d=3$ orthogonal universality class of the Anderson transition.
While the mapping in Ref. \citen{casati89} with Anderson's model of localisation is not exact, since the lattice model obtained has a quasi-periodic rather than random potential, there is clearly strong evidence that the transitions are in the same universality class.
Moreover, the mapping makes clear that the dimensionality of the lattice model is determined by the number of incommensurate frequencies.
Higher dimensional Anderson transitions could be realised in the same rotor but subject to modulation by appropriate numbers of incommensurate frequencies.
Numerical estimates for the critical exponents for Anderson's model of localisation in $d=4, 5$ and $6$ dimensions have been reported in several previous papers \cite{schreiber96, travenec02, garcia07}, with a precision approaching about $5\%$.
For $d=4$ the transition has also been studied numerically in the kicked rotor.\cite{borgonovi97,shepelyansky11}
Here, we report improved numerical estimates for $d=4$ and $d=5$ with precisions of approximately $1.2\%$ and $1.5\%$, respectively.

Another motivation for our work is the very poor agreement between the field theoretical predictions for the critical exponent and the available numerical results.
In common with other continuous phase transitions the Anderson transition is described by a field theory\cite{wegner79,schafer80,efetov83}.
One of the results of this approach is a perturbation series, in powers of $\epsilon = d - 2$, for the critical exponent.
To obtain predictions for the critical exponent from this series a Borel-Pad\'e analysis is required.
However, the analysis presented in Ref. \citen{hikami92} gives values for the critical exponent that are in clear disagreement with the available numerical results and also violate the lower bound\cite{chayes86,kramer93b} $\nu \geq 2/d$ for the critical exponent, for $d \geq 4$.
As we describe below, a new Borel-Pad\'e analysis that takes into account the asymptotic behaviour of the critical exponent in the limit $d\to\infty$ gives very much better agreement with numerical results, at least as good or better than any other approach.

The paper is organized as follows.
In Section 2 we describe the numerical estimation of the critical exponent for $d=4$ and $d=5$.
In Section 3 we describe our new Borel-Pad\'e analysis.
In Section 4 we compare the predictions for the critical exponent obtained with our new Borel-Pad\'e analysis, and also other analytical approaches, with the available numerical results.
In Section 5 we conclude.

\section{Numerical estimation of critical exponent $\nu$} \label{Numerical estimation}
\subsection{Anderson's model of localisation}
The Hamiltonian of Anderson's model of localisation is
\begin{align}
	H &= \sum_{ \mathbf{r} } \epsilon_{\mathbf{r}} |\mathbf{r} \rangle \langle \mathbf{r}| + \sum_{\mathbf{r},\mathbf{r'}} V_{\mathbf{r} \mathbf{r '}} |\mathbf{r} \rangle \langle \mathbf{r'}| \;, \label{Anderson Hamiltonian} \\
	V_{\mathbf{r} \mathbf{r'}} & =  \left \{
		\begin{array}{l}
		1 \;\;  (\mathbf{r}, \mathbf{r'} \textrm{are nearest neighbors}) \\
		0 \;\; (\textrm{otherwise})
   		\end{array} \right. \;. \label{Hopping Integral}
\end{align}
Here, $\mathbf{r},\mathbf{r'}$ are lattice points on $d$-dimensional cubic lattice.
The unit of energy has been set equal to the hopping energy.
The site energies $\epsilon_{\mathbf{r}}$ are independently and identically distributed according to the following probability density function,
\begin{eqnarray}
	p(\epsilon_{\mathbf{r}}) & = & \left \{
		\begin{array}{l}
		1/W \;\; (|\epsilon_{\mathbf{r}}| \leq W/2)\\
		\: 0 \;\;\:\, (\textrm{otherwise})
   		\end{array} \right. \;. \label{ box distribution function }
\end{eqnarray}
The choice of the probability density function does not change the universality class\cite{slevin99a,slevin14}.
The degree of disorder is set by the parameter $W$.
We used MT2203 of the Intel MKL library to generate the required random numbers.

\subsection{Transfer matrix method}

The numerical method we have used is described in detail in Ref. \citen{slevin14}. Here, we give only a very brief description and refer the reader to that reference for details.

We consider a very long (length $L_x$) quasi-one dimensional bar of cross section $L^{d-1}$ and study the localisation of electrons on this bar using the transfer matrix method.
The results of the transfer matrix calculation are an estimate of the smallest positive Lyapunov exponent $\gamma$
and the standard deviation $\sigma$ describing the statistical error in this estimate.
The smallest positive Lyapunov exponent corresponds to the inverse of the localisation length for
electrons on the bar.
The Lyapunov exponent depends on the energy $E$, the disorder $W$, the linear cross-section $L$ and the
boundary conditions imposed in the transverse direction.
We set $E=0$ throughout, which corresponds to the band centre, and we impose periodic boundary conditions in the transverse directions.

Data for $W=30,31,\cdots,40$, $L=4,6,\cdots,20$ for $d=4$, and $W=52,53, \cdots,64$, $L=4,5,\cdots,10$ for $d=5$ were accumulated.
Almost all of the data have a precision of 1\% which required between the order of $10^4$ and $10^5$ transfer matrix multiplications depending on disorder and dimension.
To avoid round-off error QR factorizations were performed after every 4 transfer matrix multiplications.
To ensure that the precision of the Lyapunov exponents was estimated correctly we set $r=4$ in Eq. (34) of Ref. \citen{slevin14}.

\subsection{Finite size scaling analysis}
To estimate the critical exponent and other quantities we analyse the data for the smallest positive Lyapunov exponent by
using the method of finite size scaling to fit the disorder and size dependence of
the dimensionless quantity
\begin{equation}
	\Gamma = \gamma L \;.
\end{equation}
As above we give only a brief description here and refer the reader to Ref. \citen{slevin14} for further details.

We found that it was not necessary to consider corrections due to irrelevant variables,
instead it was sufficient to fit the data to
\begin{equation}
	\Gamma = F ( \phi(w,L) ) \;, \label{scaling relation}
\end{equation}
where
\begin{equation}\label{one-parameter scaling variable}
  \phi(w,L) = u(w) L^{1/\nu} \;,
\end{equation}
and
\begin{equation}
w = \frac{W-W_{c}}{W_{c}} \;,
\end{equation}
with $W_{c}$ the critical disorder.
The scaling function $F$ was expanded as follows
\begin{equation}\label{expanded scaling function}
  F(\phi) = \sum^{n}_{j=0}  a_{j } \;  \phi^{j} \;.
\end{equation}
Non-linearity of the relevant scaling variable was treated using the expansion
\begin{equation}\label{scaling variable}
  u(w) = \sum^{m}_{j=1} b_{j} \, w^{j} \;.
\end{equation}
The integers $m$ and $n$ define the order of the expansions.

Considering the localised ($w>0$) and extended ($w<0$) phases separately,
the scaling law Eq. (\ref{scaling relation}) may be re-written in the form
\begin{equation}
  \varGamma = F_{\pm} \left( \frac{L}{\xi(w)} \right) \; ,
\end{equation}
where the subscript $\pm$ refers to the sign of $w$ and the correlation length is given by
\begin{equation}
	\xi(w) = \xi_{\pm}\left| u(w) \right|^{-\nu} \; .
\end{equation}
The functions $F_{\pm}$ are given by
\begin{equation}
  F_{\pm}(x) = \sum^{n}_{j=0} (\pm1)^{j} a_{j} x^{j/\nu} \; .
\end{equation}
This makes it clear that $\nu$ is the correlation length critical exponent.
The absolute scale of the correlation length (given by the constants $\xi_{\pm}$) cannot be
determined in the finite size scaling analysis.

The best fit is found by minimizing the chi-squared statistic $\chi^2$ in the usual way.
The quality of the fit is assessed using the goodness of fit probability, which is determined from the
minimum value of $\chi^2$ and the number of degrees of freedom in the fit.

\begin{figure}[ht]
\begin{center}
 \includegraphics[scale=1.0]{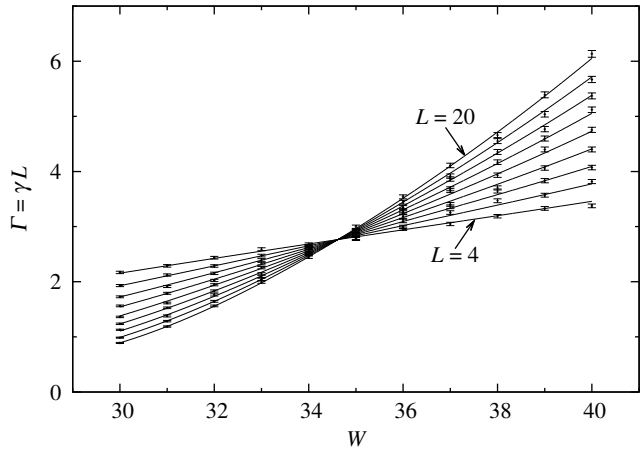}
 \caption{
 The data from the transfer matrix calculation and the finite size scaling analysis for the Anderson transition in the $d=4$ orthogonal universality class.
 Here the data are plotted versus disorder and different curves correspond to different system sizes.
 The common crossing point of different curves indicates the critical disorder separating the localised and diffusive regimes in $d=4$.
 The increase in the slope at the critical disorder with system size is related to the critical exponent.}
\label{W Gamma plot_d=4}
\end{center}
\end{figure}

\begin{figure}[ht]
\begin{center}
  \includegraphics[scale=1.0]{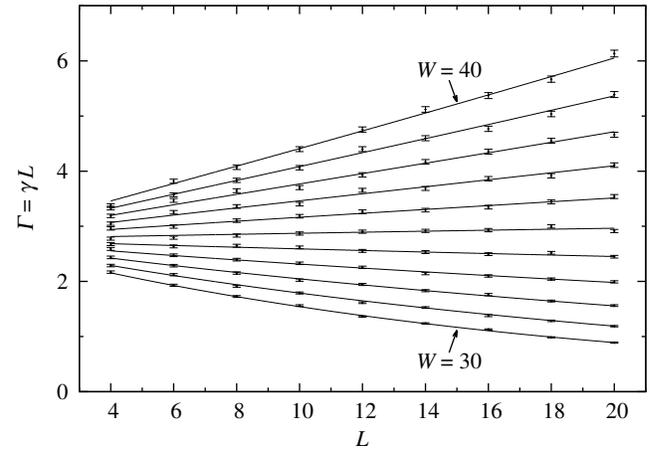}
 \caption{Here the data for the Anderson transition in the $d=4$ orthogonal universality class are plotted versus system size. The different curves correspond to different disorders.
 The change of sign of the slope indicates the transition from localised to extended states in $d=4$.}
\label{L Gamma plot_d=4}
\end{center}
\end{figure}

For $d=4$ we found $m=3$ and $n=3$ in Eqs. (\ref{expanded scaling function}) and (\ref{scaling variable})
gave an acceptable fit.
The number of parameters and the number of data used in the finite size scaling analysis were respectively $8$ and $99$.
For the best fit $\chi^{2}=91.6$ which gives a goodness of fit of $0.46$.
The fit is displayed in Figs. \ref{W Gamma plot_d=4} and \ref{L Gamma plot_d=4}.
The estimates of the critical exponent $\nu$, the critical disorder
$W_{c}$ and the critical value $\varGamma_{c}$ of $\varGamma$, together with the standard deviations of these estimates, are
\begin{align}
	\nu &= 1.156 \pm 0.014 \nonumber \;,\\
	\varGamma_{c} &=  2.76 \pm 0.01 \nonumber \;,\\
	W_{c} &=  34.62 \pm 0.03\label{nu_4d} \;.
\end{align}

\begin{figure}[ht]
\begin{center}
 \includegraphics[scale=1.0]{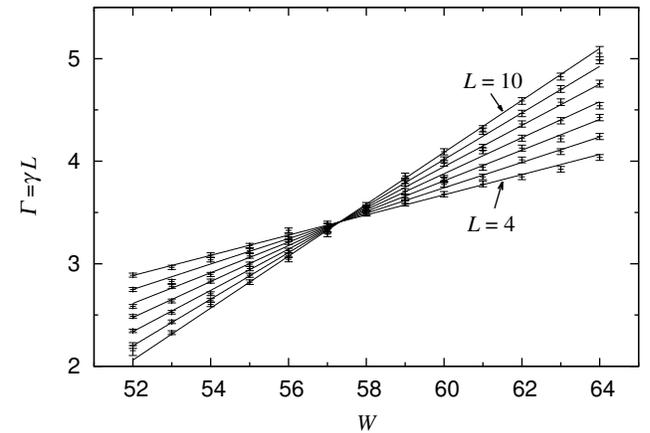}
\caption{The data from the transfer matrix calculation and the finite size scaling analysis for the Anderson transition in the $d=5$ orthogonal universality class. Compared with $d=4$ the transition occurs at a much higher disorder consistent with it becoming progressively more difficult to localize electrons in higher dimensions. }
\label{W Gamma plot_d=5}
\end{center}
\end{figure}

\begin{figure}[ht]
\begin{center}
  \includegraphics[scale=1.0]{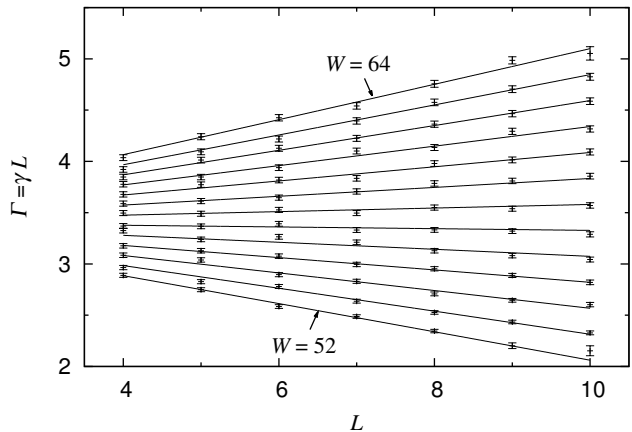}
\caption{Exactly similar to Fig. \ref{L Gamma plot_d=4} except that here the data and fit are for the Anderson transition in the $d=5$ orthogonal universality class.}
\label{L Gamma plot_d=5}
\end{center}
\end{figure}

Here, $\varGamma_{c}$ is defined by
\begin{equation}
	\varGamma_{c}=F(0) \; .
\end{equation}
For $d=5$, $m=1$ and $n=1$ gave an acceptable fit. The number of parameter and the number of data are
$4$ and $91$.
For the best fit $\chi^{2}=84.0$ and the goodness of fit is $0.57$.
The fit is displayed in Figs. \ref{W Gamma plot_d=5} and \ref{L Gamma plot_d=5}.
\begin{align}
	\nu &= 0.969 \pm 0.015 \nonumber \;, \\
	\varGamma_{c} &=  3.41 \pm 0.01 \nonumber \;, \\
	W_{c} &=  57.3 \pm 0.05 \label{nu_5d} \;.
\end{align}
These results agree with numerical estimates reported in previous works \cite{zharekeshev98,markos06,garcia07} but are considerably more precise.



\section{The Borel-Pad\'e analysis} \label{Analytical estimation}

In Ref. \citen{hikami92} a perturbation analysis carried to five-loop order yielded the following result for
dimensionality dependence of the critical exponent
\begin{align}
	\nu=\frac{1}{\epsilon}-\frac{9}{4}\zeta(3)\epsilon^{2}+\frac{27}{16}\zeta(4)\epsilon^{3}+O(\epsilon^{4}) \;. \label{asymptotic_zero}
\end{align}
Here
\begin{equation}
  \epsilon \equiv d -2 \;.
\end{equation}
Straightforward application of Borel-Pad\'e analysis\cite{bender99} for this expression gives the following
\cite{hikami92}
\begin{align}
	\nu\simeq\frac{1}{\epsilon^{2}}\int_{0}^{\infty}dt e^{-t/\epsilon}\frac{ 1+\frac{3\zeta(4)}{16\zeta(3)}t }{ 1+\frac{3\zeta(4)}{16\zeta(3)}t+\frac{3}{8}\zeta(3)t^{3} } \;.
\label{BPhikami}
\end{align}
The values for the exponents obtained from this expression for the $d=3,4,5$ and $6$ dimensional orthogonal universality classes are listed in Table \ref{table:exponents}.
There is clear discrepancy with the numerical results.
As we now describe, the agreement is considerably improved if the Borel-Pad\'e analysis is revised to 
incorporate the asymptotic behaviour for $\epsilon \to \infty$.

On the basis of several approaches\cite{efetov90,mirlin94,schreiber96,garcia08}, it is widely thought (though not universally\cite{suslov14}) that  $d=\infty$ is the
upper critical dimension and that
\begin{equation} \label{asymptotic_infinity}
  \nu(\epsilon) \sim \frac{1}{2} \; \left( \epsilon \to \infty \right) \;.
\end{equation}
However, for Eq. (\ref{BPhikami}) we have
\begin{equation}
  \lim_{\epsilon \to \infty} \nu(\epsilon) = 0 \;.
\end{equation}
To correct this we rewrite the
$\epsilon$-expansion as
\begin{equation}
  \nu = \frac{1}{2} + \frac{1}{\epsilon} f(\epsilon) \;,
\end{equation}
with
\begin{equation}
  f(\epsilon) = 1 -\frac{1}{2}\epsilon-\frac{9}{4}\zeta(3)\epsilon^{3}+\frac{27}{16}\zeta(4)\epsilon^{4}+O(\epsilon^{5})
\end{equation}
After this separation, the Borel-Pad\'e analysis is applied only to $f(\epsilon)$,
\begin{equation}
	f(\epsilon) \simeq \frac{1}{\epsilon} \mathcal{P}\int_{0}^{\infty}dt e^{-t/\epsilon} h(t) \;.
\end{equation}
Here, $\mathcal{P}$ indicates the Cauchy principal value, and
\begin{equation}
h(t) = \frac{ 1 +\left(\frac{3\zeta(4)}{16\zeta(3)}-\frac{1}{2}\right)t -\left(\frac{3}{4}\zeta(3)+\frac{3\zeta(4)}{32\zeta(3)}\right)t^{2} }{ 1+\frac{3\zeta(4)}{16\zeta(3)}t-\frac{3}{4}\zeta(3)t^{2} }\;.
\end{equation}
From this we obtain the following expression
\begin{equation}\label{BPnew}
  \nu \simeq\frac{1}{2}+\frac{1+\frac{\zeta(4)}{8\zeta(3)^2}}{\epsilon}-\frac{1}{3\zeta(3)\epsilon^{2}} g(\epsilon) \;,
\end{equation}
where
\begin{equation} \label{function g}
  g(\epsilon) = c_{+}e^{-t_{+}/\epsilon}\mathrm{Ei}\left(\frac{t_{+}}{\epsilon}\right)+c_{-}e^{-t_{-}/\epsilon}\mathrm{Ei}\left(\frac{t_{-}}{\epsilon}\right)\;.
\end{equation}
Here, $c_{\pm}, t_{\pm}$ are given by
\begin{align}
	c_{\pm}&=1+\frac{3\zeta(4)^2}{64\zeta(3)^3}\pm\frac{9\zeta(4)}{\sqrt{768\zeta(3)^3+9\zeta(4)^2}} \left( 1+\frac{\zeta(4)^{2}}{64\zeta(3)^3} \right) \;,  \nonumber \\
      	          &\simeq 1.3, 0.7632 \;. \nonumber \\
	t_{\pm}&=\frac{3\zeta(4)\pm\sqrt{768\zeta(3)^3+9\zeta(4)^2}}{24\zeta(3)^2} \;, \nonumber \\
	          &\simeq 1.151, -0.9637 \;.
\end{align}
In the Appendices we confirm that this new Borel-Pad\'e analysis has the required asymptotic behaviour, i.e. Eqs. (\ref{asymptotic_zero}) and (\ref{asymptotic_infinity})
in the respective limits

\section{Discussion}

The main limitation of the epsilon expansion method is that it is only practicable to calculate a
limited number of terms in the expansion.
The predications for the critical exponent are, therefore, necessarily approximate.
Numerical simulations are subject to limitations of systems size and computer time and,
therefore, also lead to results that are approximate.
An important difference is that for numerical results it is possible to reliably specify their precision.
In the following, we compare the various analytical results for the exponents with
the available numerical results.

The self-consistent theory\cite{vollhardt92} of Anderson localisation predicts the following dimensionality dependence of the critical exponent
\begin{eqnarray}\label{selfconsistent}
  \nu &=& \frac{1}{\epsilon} \; \; \; (2<d<4) \;, \nonumber \\
  \nu &=& \frac{1}{2} \; \; \; (d \geq 4) \;.
\end{eqnarray}
Reference to Table \ref{table:exponents} shows immediately that, as expected, the predictions of the self-consistent theory for critical phenomena are not quantitatively accurate.
Moreover, the numerical results leave no doubt that the dimensionality dependence of the exponent persists beyond $d=4$.
The prediction that the upper critical dimension
is $d=4$ is, therefore, not correct.

\begin{figure}[ht]
\begin{center}
 \includegraphics[scale=1.0]{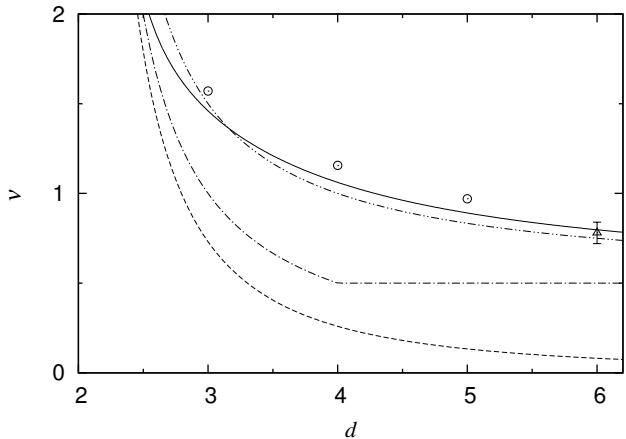}
\caption{The dimensionality dependence of the critical exponent $\nu$ of the Anderson transition.
The points are numerical estimates in $d=3, 4, 5$ (($\circ$) Ref. \citen{slevin14} and this work) and $d=6$ (($\triangle$) Ref. \citen{garcia07}).
Error bars are standard deviations.
They are omitted when the error is smaller than the symbol size.
The lines are analytical predictions:
our new Borel-Pad\'e analysis Eq. (\ref{BPnew}) (solid),
the semiclassical theory Eq. (\ref{semiclassical}) (dash dot dot),
the self-consistent theory Eq. (\ref{selfconsistent}) (dash dot), and
the Borel-Pad\'e analysis of Eq. (\ref{BPhikami}) (dash). }
\label{plotepsa}
\end{center}
\end{figure}

Reference to Table \ref{table:exponents} and Fig. \ref{plotepsa} shows that the predictions of the Borel-Pad\'e analysis of Ref. \citen{hikami92} given by Eq. (\ref{BPhikami}) are in poor agreement with the numerical results.
The asymptotic behaviour for $d\to\infty$ is clearly incorrect.
Moreover, the estimates for higher dimensions violate the well known lower bound\cite{chayes86,kramer93b} for the exponent
\begin{equation}\label{lowerbound}
  \nu \geq \frac{2}{d}\;.
\end{equation}

According to the semi-classical theory of the Anderson transition presented in Ref. \citen{garcia08}
\begin{equation}\label{semiclassical}
  \nu = \frac{1}{2} + \frac{1}{\epsilon} \; \; \; (d>2) \;,
\end{equation}
The asymptotic behaviour for $\epsilon \to \infty$ agrees with Eq. (\ref{asymptotic_infinity}).
However, the asymptotic behaviour for $\epsilon \to 0$ is only correct at leading order.
Nevertheless, reference to Table \ref{table:exponents} and Fig. \ref{plotepsa} shows that the
agreement with the numerical results for $d= 3, 4,5$ and $6$, though certainly not exact, is much better than either the original Borel-Pad\'e analysis Eq. \ref{BPhikami} or the self-consistent theory.

\begin{figure}[ht]
\begin{center}
\includegraphics[scale=1.0]{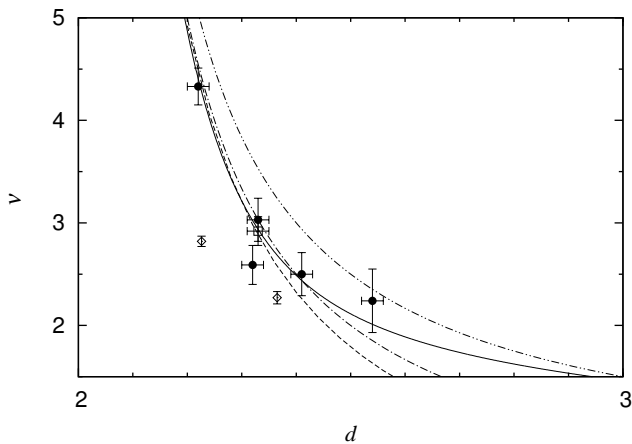}
\caption{The dimensionality dependence of the critical exponent $\nu$ of the Anderson transition for various fractals.
The points are numerical estimates (($\Box$) Ref. \citen{song97}, ($\Diamond$) Ref. \citen{travenec02}, {($\bullet$) Ref. \citen{schreiber96}}) ).
The lines have the same meaning as in Fig. \ref{plotepsa}.}
\label{plotepsb}
\end{center}
\end{figure}

Finally, we turn to our new Borel-Pad\'e analysis.
Again by reference to Table \ref{table:exponents} and Fig. \ref{plotepsa} we see that the agreement with the numerical results is slightly worse for $d=3$ but better for $d=4,5$ and 6
when compared with the semi-classical theory,

So far we have considered only integer dimensions.
Numerical results for the Anderson transition on fractals are also available\cite{schreiber96,song97,travenec02}
(See Table \ref{table:fractals} and Fig. \ref{plotepsb}).
Fractals are described by several dimensions, among them the spectral dimension.
Previous numerical work\cite{schreiber96} has established that it is the spectral dimension $d_s$ that determines the universality class.
Therefore, we identify $\epsilon$ with $d_s-2$ for fractals.
The available results allow us to focus on the interval near $d=2$ where the epsilon expansion should be most accurate.
(Unfortunately, the authors of Refs. \citen{schreiber96} and \citen{travenec02} do not agree on the spectral dimension of some fractals.)
For sufficiently small $\epsilon$ both the original and the new Bore-Pad\'e analysis, as well as the self-consistent theory, are in good agrement with the numerical data.
The limitations of the semi-classical theory, however, become clearly apparent.

\begin{table}[ht]
\begin{tabular}{|c|c|c|c|c|} \hline
		    & $d=3$ & $d=4$ & $d=5$ & $d=6$ \\ \hline \hline
Eq. (\ref{selfconsistent})   &   1   & 0.5 & 0.5 & 0.5 \\
Eq. (\ref{BPhikami}) & $0.731$ & $0.259$  & $0.133$ & $0.081$   \\
Eq. (\ref{semiclassical}) & $1.5$ & $1.0$  & $0.83$ & $0.75$   \\
Eq. (\ref{BPnew}) & $1.460$ & $1.061$  & $0.891$ & $0.798$   \\ \hline
This study&  				     & $1.156\pm.014$    &	$0.969\pm.015$  	&			\\
ref. \citen{slevin14} &$1.571\pm.004$&	  	  		   &		 		      &			\\
ref. \citen{markos06}  &				     & $1.12\pm.05$  & $0.94\pm.05$  &			\\
ref. \citen{rodriguez11}  &	$1.590\pm.006$			     & &	  　　　                &			\\
ref. \citen{garcia07}  &$1.52\pm.06$& $1.03\pm.07$&	$0.84\pm.06$  & $0.78\pm.06$\\
ref. \citen{schreiber96} & $1.35\pm.15$ & $1.03\pm.17$ & & \\
refs. \citen{zharekeshev98} and \citen{zharekeshev95} & $1.45\pm.08$ & $1.1\pm.2$  & & \\ \hline
\end{tabular}
\caption{Comparison between numerical and analytical estimates of the  critical exponent $\nu$ for $d=3,4,5,6$. }
\label{table:exponents}
\end{table}

\begin{table}[ht]
\begin{tabular}{|l|c|c|c|c|} \hline
 & Eq. (\ref{BPnew}) & ref. \citen{schreiber96} & ref. \citen{travenec02} & ref. \citen{song97} \\ \hline \hline
$d_s=2.22$   & $4.41$ & $4.33\pm.18$  &			    & \\
$d_s=2.226$ & $4.28$ & 			 & $2.82\pm.05$ & \\
$d_s=2.32$   & $3.02$ & $2.59\pm.19$ &  			    & \\
$d_s=2.33$   & $2.94$ & $3.03\pm.21$ &  			    & $2.92\pm.14$ \\
$d_s=2.365$ & $2.68$ & 			 & $2.27\pm.06$ & \\
$d_s=2.41$  & $2.44$  & $2.50\pm.21$ &  			    & \\
$d_s=2.54$  & $2.01$  & $2.24\pm.31$  &  			    & \\ \hline
\end{tabular}
\caption{Comparison between numerical and analytical estimates of the  critical exponent $\nu$ for $d=2 \sim 3$. }
\label{table:fractals}
\end{table}

\section{Conclusions}

We have presented more precise numerical estimates for the critical exponent of the Anderson transition in the $d=4$ and $d=5$ orthogonal universality classes.
It may be possible to test these predictions in future experiments on the quantum kicked rotor system in cold atomic gases.

We have also presented a new  Borel-Pad\'e analysis, which takes account of the asymptotic behaviour of the exponent
as $d\to \infty$.
This new Borel-Pad\'e analysis gives very reasonable agreement with the available numerical estimates of the exponent.
The agreement could be improved if either more terms in the $\epsilon$-expansion Eq. (\ref{asymptotic_zero}), 
or alternatively more terms in the asymptotic expansion for $\epsilon \to \infty$ beyond the leading order of Eq. (\ref{asymptotic_infinity}), were known.
Such information could be incorporated in a further revision of the Bore-Pad\'e analysis.

Another possibility would be to incorporate information about the asymptotic behaviour of the coefficients in 
the $\epsilon$-expansion for large order along the lines of Refs. \citen{pelissetto02} and \citen{guillou77}.

\section*{Acknowledgements}

We would like to thank Michael Schreiber for supplying the numerical data used in Ref. \citen{schreiber96}.
We would also like to thank Tomi Ohtsuki for a critical reading of the manuscript.

\appendix

\section{Asymptotic behaviour for $\epsilon \to 0$}
Here we confirm that the new Borel-Pad\'e analysis has the correct asymptotic behaviour for
$\epsilon \to 0$, i.e., Eq. (\ref{asymptotic_zero}).

The function $f$ is said to be asymptotic to the function $g$  as $x \to x_0$ if\cite{bender99}
\begin{equation}
  \lim_{x \to x_0} \frac{f(x)}{g(x)} = 1 \; .
\end{equation}
This is usually written
\begin{equation}
  f(x) \sim g(x) \; \left( x \to x_0 \right) \; .
\end{equation}
A series
\begin{equation}
  \sum_{n=0}^{\infty} a_n \left( x - x_0 \right)^n \; ,
\end{equation}
is said to be asymptotic to the function $f(x)$ if\cite{bender99}
\begin{equation}
  f(x) - \sum_{n=0}^{N} a_n \left( x - x_0 \right)^n \sim a_M \left( x - x_0 \right)^M \; \left( x \to x_0 \right) \;,
\end{equation}
where $a_M$ is the first non-zero coefficient after $a_N$.

The exponential integral Ei$(x)$ is defined as\cite{jeffrey04}
\begin{equation}
  {\rm Ei}(x) = \mathcal{P} \int_{-\infty}^{x} \frac{e^t}{t} {\rm d}t \;,
\end{equation}
and has the following asymptotic expansion for $|x| \to \infty$
\begin{equation}
	\mathrm{Ei}(x) \sim \mathrm{e}^{x} \sum^{\infty}_{n=1} \frac{(n-1) !}{x^{n}} \; .
\end{equation}
Using this expansion in Eq.(\ref{function g}) we find for $\epsilon \to 0^+$,
\begin{equation}\label{AsymptoticExpansionZero}
  g(\epsilon) \sim \sum^{\infty}_{n=1} g_{n} \epsilon^{n} \;,
\end{equation}
with
\begin{equation}\label{gn}
  g_{n} = (n-1)! \left( \frac{c_{+}}{t_{+}^{\, n}} +  \frac{c_{-}}{t_{-}^{\, n}}  \right) \;.
\end{equation}
For $n=1, 2$ we have explicitly
\begin{align}
	g_{1} = \frac{3\zeta(4)}{8\zeta(3)} , \; \; \; g_{2} = \frac{3\zeta(3)}{2} \; .
\end{align}
Eq. (\ref{gn}) is the solution of a three term recurrence relation and we can use this fact to calculate $g_{n}$ for $n \ge 3$,
\begin{align}
	\frac{g_{n}}{(n-1)!} &= \left( \frac{t_{+} + t_{-}}{t_{+}t_{-}} \right) \frac{g_{n-1}}{(n-2)!} - \left( \frac{1}{t_{+}t_{-}} \right) \frac{g_{n-2}}{(n-3)!} \nonumber \\
	\iff g_{n} &= - \frac{3\zeta(4)}{16\zeta(3)} (n-1) g_{n-1} + \frac{3\zeta(3)}{4} (n-1)(n-2)g_{n-2} \; .
\end{align}
Using this formula, higher coefficients are more easily calculated,
\begin{align}
	g_{3} &= 0 \nonumber \; , \nonumber \\
	g_{4} &= \frac{27\zeta(3)^2}{4} \; , \nonumber \\
	g_{5} &= - \frac{81\zeta(3)\zeta(4)}{16} \; .
\end{align}
Upon substitution in to Eq. (\ref{BPnew}), we find agreement term by term with Eq. (\ref{asymptotic_zero}).

\section{Asymptotic behaviour for $\epsilon \to \infty$}
Here we demonstrate that the new Borel-Pad\'e analysis has the required asymptotic behaviour for
$\epsilon \to \infty$, i.e., Eq. (\ref{asymptotic_infinity}).

The exponential integral has the following asymptotic expansion for $x \to 0$,\cite{jeffrey04}
\begin{equation}
	\mathrm{Ei}(x) \sim \gamma + \ln|x| + \sum_{n=1}^{\infty} \frac{x^{n}}{n \cdot n!} \;,
\end{equation}
where $\gamma$ is Euler-Mascheroni constant.
After substitution in Eq.(\ref{function g}) we obtain
\begin{align}
	  g(\epsilon) &\sim c_{+}e^{-t_{+}/\epsilon} \left(\gamma + \ln \left| \frac{t_{+}}{\epsilon} \right|
			+ \sum_{n=1}^{\infty} \frac{t_{+}^{\, n}}{n \cdot n!} \epsilon^{-n} \right) \nonumber \\
		&+ c_{-}e^{-t_{-}/\epsilon} \left(\gamma + \ln \left| \frac{t_{-}}{\epsilon} \right|
			+ \sum_{n=1}^{\infty} \frac{t_{-}^{\, n}}{n \cdot n!} \epsilon^{-n} \right) \; .
\end{align}
It follows that
\begin{equation}
	\lim_{\epsilon \to \infty} \frac{g(\epsilon)}{\epsilon^{2}} = 0 \: ,
\end{equation}
which leads to Eq. (\ref{asymptotic_infinity}).

\bibliography{references}

\end{document}